# Bulk superconductivity in pressurized trilayer nickelate $Pr_4Ni_3O_{10}$ single crystals

Enkang Zhang[1,2,†], Di Peng[3,†], Yinghao Zhu[1,†], Lixing Chen[1], Bingkun Cui[1], Xingya Wang[4], Wenbin Wang[5], Qiaoshi Zeng[3,6,*], Jun Zhao[1,2,5,7,*]

[1]*State Key Laboratory of Surface Physics and Department of Physics, Fudan University, Shanghai 200433, China*

[2]*Shanghai Research Center for Quantum Sciences, Shanghai 201315, China*

[3]*Shanghai Key Laboratory of Material Frontiers Research in Extreme Environments (MFree), Shanghai Advanced Research in Physical Sciences (SHARPS), Shanghai 201203, China*

[4]*Shanghai Synchrotron Radiation Facility, Shanghai Advanced Research Institute, Chinese Academy of Sciences, Shanghai 201204, China*

[5]*Institute of Nanoelectronics and Quantum Computing, Fudan University, Shanghai 200433, China*

[6]*Center for High Pressure Science and Technology Advanced Research, Shanghai 201203, China*

[7]*Hefei National Laboratory, Hefei 230088, China*

## Abstract

**The discovery of superconductivity in pressurized bilayer and trilayer nickelates has generated significant interest. However, their superconducting properties are often dependent on sample quality and pressure conditions, complicating the interpretation of the underlying physics. Finding new systems with optimized bulk superconducting properties is therefore important for advancing our understanding of these materials. Unlike cuprates, where trilayer compounds typically exhibit the highest transition temperature ($T_c$), the bilayer nickelate $La_3Ni_2O_7$ has thus far outperformed the trilayer $La_4Ni_3O_{10}$ in reported $T_c$. Whether the trilayer nickelates have achieved the optimal $T_c$ remains unclear, with various scenarios suggesting different possibilities. Here, we report the discovery of bulk superconductivity in pressurized $Pr_4Ni_3O_{10}$ single crystals,**

**achieving a maximum onset $T_c$ of 40.5 K at 80.1 GPa, significantly exceeding the 30 K observed in $La_4Ni_3O_{10}$. The bulk nature of superconductivity is confirmed by zero resistance and a strong diamagnetic response below $T_c$ with a superconducting volume fraction exceeding 80%. These findings establish trilayer nickelates as genuine bulk high-temperature superconductors, provide new insights into the mechanisms driving superconductivity, and point to a promising route toward further enhancing superconducting properties in nickelates.**

I. Introduction

High-temperature (high-$T_c$) superconductivity has fascinated condensed matter physicists and materials scientists, offering both technological applications and fundamental insights into strongly correlated electron systems[1-3]. Since the discovery of high-$T_c$ cuprates more than three decades ago, the search for new families of superconductors has remained a central focus in this field[4-6].

Nickel-based materials have long been proposed as promising candidates for replicating the high-temperature superconductivity observed in cuprates[7]. This interest led to the discovery of superconductivity in infinite-layer nickelates[8], which share important structural and electronic similarities with cuprates. Recently, Ruddlesden-Popper (RP) phase bilayer nickelate $La_3Ni_2O_7$ was shown to exhibit superconductivity under high pressures, with transition temperatures ($T_c$) approaching 80 K[9-11]. This unexpected finding prompted discussions about the underlying mechanisms of superconductivity, including analogies to cuprates, and the potential for multi-orbital physics that goes beyond simple cuprate-like models[12].

Despite these advances, challenges remain in fully understanding the superconducting properties of nickelates under pressure. Notably, superconductivity in these materials is often sensitive to sample quality and pressure conditions, and the superconducting volume fractions are frequently low[13-16]. These sample-dependent variations hinder the identification of the precise mechanisms governing superconductivity in nickelates and raise fundamental questions about the nature of the superconducting phase in these materials.

Very recently, studies have revealed that trilayer nickelate $La_4Ni_3O_{10}$ single crystals host bulk superconductivity with rather high superconducting volume fractions of around 86% under pressure[17]. Drawing from analogies with cuprates, where trilayer systems often achieve the highest $T_c$[18-20], it was anticipated that $La_4Ni_3O_{10}$ might exhibit an even higher $T_c$ than $La_3Ni_2O_7$. However, experimental reports on pressurized $La_4Ni_3O_{10}$ indicate a $T_c$ ranging from 20 to 30 K[17, 21-23], substantially below the 80 K observed for $La_3Ni_2O_7$. These observations raise fundamental questions about whether trilayer nickelates have yet reached their intrinsic superconducting potential or if additional tuning strategies are necessary to optimize their electronic and structural properties.

A promising strategy for tuning superconductivity in RP phase nickelates involves inducing "internal chemical pressure" through chemical substitution of rare-earth ions[24, 25]. These substitutions may alter lattice constants, bond lengths or angles, thereby modifying the electronic environment without introducing extra disorder on the $NiO_2$ plane. By introducing internal chemical pressure, in combination with applied hydrostatic pressure, it may be possible to explore superconductivity in new dimensions in nickelates.

## II. Results

In this work, we investigate the trilayer nickelate $Pr_4Ni_3O_{10}$ single crystals under ambient and high-pressure conditions. Polycrystalline samples were synthesized using conventional solid-state methods, followed by pressing the material into precursor rods and sintering them in an air atmosphere. High-quality single crystals of $Pr_4Ni_3O_{10}$ were subsequently grown from the sintered rods using a high-pressure vertical optical floating-zone furnace (see Sec. V). X-ray diffraction (XRD) measurements on cleaved surfaces of $Pr_4Ni_3O_{10}$ crystals revealed sharp (0 0 $L$) Bragg peaks, consistent with a trilayer structure (Supplementary Fig. 1a). Further Rietveld refinements of X-ray diffraction data from powdered single crystals confirmed a trilayer structure similar to that of $La_4Ni_3O_{10}$, but with reduced lattice parameters (Supplementary Fig. 1b, Supplementary Table 1). This reduction suggests the influence of internal chemical pressure resulting from the smaller ionic radius of $Pr^{3+}$ compared to $La^{3+}$. The refined crystal structure belongs to the monoclinic $P2_1/a$ space group [Figs. 1(a) and 1(b)], consistent with previous studies[26, 27].

To explore the structural evolution of $Pr_4Ni_3O_{10}$ under pressure, we conducted synchrotron X-ray diffraction measurements on powdered single crystals. At low pressures, the data confirm a monoclinic $P2_1/a$ structure [Figs. 1(a) and 1(b)]. Upon increasing the pressure above 35 GPa, a structural transition to the tetragonal $I4/mmm$ phase is observed, as evidenced by the merging of the (020) and (200) diffraction peaks [Figs. 1(c), 1(f) and 1(g)]. These findings are analogous to those reported for $La_4Ni_3O_{10}$, although $Pr_4Ni_3O_{10}$ exhibits a notably higher transition pressure[17]. Our refinements reveal that the Ni–O–Ni bonding angle between adjacent $NiO_2$ layers increases from 161.4° to 180° during this phase transition [Figs. 1(a), 1(d), and 1(e)]. This change indicates that the enhancement of interlayer coupling under pressure is likely a universal feature of trilayer nickelates.

Thermodynamic measurements were also conducted at ambient pressure. The heat capacity data exhibit a sharp peak at $T_d = 157$ K, indicating the onset of density wave ordering [Fig. 2(a)], which involves an intertwined spin and charge density wave[28]. In addition to the density wave transition at 157 K, the heat capacity data also show an anomaly around 5 K, likely corresponding to the development of magnetic order of the Pr moments (Supplementary Fig. 2)[26, 27].

Concurrently, magnetic susceptibility measurements exhibit a subtle anomaly at $T_d = 157$ K [Fig. 2(b)], which can be seen more clearly in the derivative of the susceptibility plot. Notably, the susceptibility anomaly at $T_d$ is much weaker than that observed in $La_4Ni_3O_{10}$[17], which is likely because the transition is masked by the dominated paramagnetic susceptibility from $Pr^{3+}$ local moments.

To explore the transport properties under pressure, resistance measurements were conducted up to 80.1 GPa using diamond anvil cells (DAC). Because of the extreme sensitivity of superconducting properties to pressure conditions in ceramic-like materials such as nickelates, cuprates and chromium arsenides[10, 17, 29], helium was utilized as the pressure-transmitting medium to ensure ideal hydrostatic conditions. At ambient pressure, the resistance of $Pr_4Ni_3O_{10}$ exhibits metallic behavior with a sharp kink corresponding to $T_d = 157$ K [Fig. 2(c)]. Upon increasing the pressure to 33.1 GPa, the anomaly associated with $T_d$ becomes weakened and shifts to 89 K, while a pronounced decrease in resistance emerges below approximately 25 K, signaling the onset of a superconducting transition [Fig. 3(a)]. As the pressure is further increased, the density-wave order diminishes and $T_c$ continues to rise, achieving zero resistance above 48.9 GPa (Fig. 3a, 3b). Ultimately, at a pressure of 80.1 GPa, the superconducting onset temperature reaches around 40.5 K [Fig. 3(a)], significantly higher than the 20-30 K reported for $La_4Ni_3O_{10}$[17].

To clarify the evolution of the density-wave order and superconductivity under pressure in more detail, we carefully prepared another sample in a helium DAC with optimized hydrostaticity. This careful loading allows for a clearer observation of the density-wave and superconducting transitions, showing an onset superconducting transition $T_c$ of 40.2 K and a zero-resistance $T_c$ of 24 K at 56.7 GPa (Fig. 4), both significantly higher than those in $La_4Ni_3O_{10}$ [17]. These results confirm that the observed effects are reproducible.

Interestingly, the normal-state resistance of $Pr_4Ni_3O_{10}$ deviates from conventional metallic behavior, following a power-law relationship $\rho(T) = \rho_0 + AT^n$ with an exponent $n$ ranging from 1 to 1.5, depending on the pressure and temperature range fitted [Figs. 3(a) and 4(a)]. An exponent smaller than 2 suggests non-Fermi liquid behavior, potentially induced by strong electron correlations or spin fluctuations near a quantum critical point. This behavior is somewhat different from that of $La_4Ni_3O_{10}$, where a linear temperature dependence of resistance is observed in the normal state near the pressure range of optimal $T_c$[17]. The difference between $La_4Ni_3O_{10}$ and $Pr_4Ni_3O_{10}$ may be related to the influence of the $Pr^{3+}$ local moments. Future studies should explore this discrepancy further.

Figure 3(c) illustrates the effects of an external magnetic field on the temperature-dependent resistance of $Pr_4Ni_3O_{10}$. Under an applied magnetic field perpendicular to the *ab* plane, the superconducting transition is progressively suppressed, which provides further confirmation that the observed transition is indeed due to superconductivity. The suppression of superconductivity with increasing magnetic field can be described by the Ginzburg-Landau (GL) theory, using the relationship:

$$H_{c2}(T) = H_{c2}(0)[(1-t^2)/(1+t^2)],$$

where $t = T/T_c$ is the reduced temperature, and $T_c$ is defined as the temperature at which the resistance reaches 90% of its normal state value near the onset of superconductivity. By fitting the experimental data to this GL form, we obtained an upper critical field $H_{c2}$ of approximately 53 T at 75.1 GPa [Fig. 3(d)]. This value exceeds the upper critical fields reported for $La_4Ni_3O_{10}$, highlighting the enhanced superconducting robustness of $Pr_4Ni_3O_{10}$ under high pressure. Moreover, our calculation of the in-plane superconducting coherence length for $Pr_4Ni_3O_{10}$ at 75.1 GPa is about 25 Å. The short superconducting coherence length suggests that the Cooper pairs are spatially confined, pointing to strong electron pairing interactions similar to those found in other high-temperature superconductors[2, 3, 30].

To further substantiate the intrinsic nature of superconductivity in $Pr_4Ni_3O_{10}$, we performed dc magnetic susceptibility measurements under high pressures using a custom-built Be-Cu alloy miniature DAC. Neon was employed as the pressure-transmitting medium to ensure optimal hydrostatic conditions. In the raw zero-field-cooled (ZFC) susceptibility data at 40.2 GPa, strong diamagnetic signals were clearly observed below $T_c$, confirming the presence of the Meissner effect [Fig. 5(a)]. When the pressure was increased to 49.5 GPa, the upper limit achievable in our miniature cell, $T_c$ rose [Fig. 5(b)]. This pressure-induced increase in $T_c$ is consistent with trends observed in resistance measurements. Together, these susceptibility and resistance data unambiguously confirm that the observed transition is due to superconductivity.

Furthermore, analysis of the DC magnetic susceptibility data revealed superconducting volume fractions of approximately 85% at 40.2 GPa and 88% at 49.5 GPa, after correcting for the demagnetizing factor associated with the sample geometry (see Sec. V, Supplementary Fig. 3). These substantial superconducting volume fractions demonstrate that superconductivity is an intrinsic bulk property of the $Pr_4Ni_3O_{10}$ phase, rather than arising from minor secondary phases or filamentary regions.

Figure 5(c) summarizes the structural evolution, density wave order, and superconductivity phase diagram of $Pr_4Ni_3O_{10}$ under pressure. The results reveal that pressure suppresses the density wave order and induces superconductivity, a behavior analogous to other high-temperature superconductors. Additionally, a monoclinic-to-tetragonal phase transition is observed at around 35 GPa, which is notably higher than that (13-15 GPa) observed in $La_4Ni_3O_{10}$[17].

### III. Discussion

The markedly higher $T_c$ observed in $Pr_4Ni_3O_{10}$ compared to $La_4Ni_3O_{10}$ provides important insight into the mechanism driving nickelate superconductivity. Unlike conventional doping, substituting $Pr^{3+}$ for $La^{3+}$ does not induce extra carrier doping or disorder; rather, the smaller ionic radius of $Pr^{3+}$ generates internal chemical pressure, reducing the lattice constants and the $c/a$ ratio[17]. This lattice-parameter modification appears to enhance exchange couplings, which has been proposed by theory as the primary driver of electron pairing in nickelates[25, 31]. In particular, these theoretical studies suggest that substituting smaller rare-earth ions strengthens interlayer exchange coupling through structural modifications, thereby boosting the pairing energy and ultimately increasing $T_c$, which is consistent with our experiments. Collectively, these observations point to a strong interplay between structural tuning and superconductivity in nickelates, highlighting the idea that targeted chemical substitutions aimed at enhancing exchange coupling can be an effective strategy for improving superconductivity.

Furthermore, the reduced ionic radius of $Pr^{3+}$ could increase the three dimensionality of the electronic structure. In cuprates, the enhanced three dimensionality may not be helpful for superconductivity as it reduces quantum fluctuations. However, in nickelates, the situation may differ. It has been proposed that interlayer coupling involving $d_{z^2}$ bands, which are inherently more three dimensional than $d_{x^2-y^2}$ bands[32-35], plays a crucial role in the pairing mechanism[31, 36-40]. Consequently, increased three-dimensionality and enhanced interlayer coupling could be favorable for superconductivity in nickelates. This idea aligns with the observed monoclinic-to-tetragonal phase transition, which enhances interlayer coupling and is accompanied by the appearance of superconductivity in this system. Indeed, it has been shown that $Pr_4Ni_3O_{10}$ exhibits rather weak anisotropy between in-plane and out-of-plane resistances[26].

These analyses suggest that similar trends may apply to other nickelate superconductors and offer an effective pathway for enhancing superconductivity by substituting smaller rare-earth elements, such as Nd, Sm, and Gd, in both bilayer and trilayer nickelates. However, it should be noted that when extremely large internal pressures are applied, the increased orbital overlap and hopping between nearest neighbors become more pronounced, potentially broadening the electronic bandwidth, which could reduce electronic correlations and weaken pairing strength[24]. Therefore, an optimal superconducting regime likely exists, where exchange couplings and electron correlations are finely balanced. Following this approach, by combining rare-earth doping with external pressure, and possibly with substrate strain[41, 42], further enhancement of superconductivity in nickelates could be achieved, potentially even under ambient or low pressures.

## IV. Conclusion

In summary, our experiments demonstrate that $Pr_4Ni_3O_{10}$ exhibits robust bulk superconductivity under high pressure, achieving transition temperatures up to 40.5 K at 80.1 GPa, which is substantially higher than the 30 K observed in its La-based counterpart. This significant enhancement underscores the critical role of internal chemical pressure, induced by the smaller ionic radius of $Pr^{3+}$, in enhancing magnetic interactions essential for high-temperature superconductivity. Moreover, the substantial superconducting volume fraction of approximately 88% affirms that superconductivity in $Pr_4Ni_3O_{10}$ is an intrinsic bulk property, rather than contributions from secondary phases or filamentary regions. The 30% increase in $T_c$ through Pr substitution suggests that further chemical substitutions with smaller rare-earth ions could open the door to discovering superconductors with even higher $T_c$ values, in both trilayer and bilayer configurations, thereby pushing the boundaries of superconducting performance in nickelates.

## V. Methods

**Growth of $Pr_4Ni_3O_{10}$ single crystals**

The synthesis of the precursor powder for the $Pr_4Ni_3O_{10}$ compound was conducted using a conventional solid-state reaction approach. $Pr_6O_{11}$ (Aladdin, 99.99%) and NiO (Aladdin, 99.99%) were used as raw materials, with an additional 0.5% NiO added to compensate for potential volatilization during crystal growth. Before weighing, the $Pr_6O_{11}$ powder was dried to eliminate moisture absorbed during storage. Subsequently, the mixture was finely ground and mixed, followed by calcination at 1373 K for 24 hours, with two repeated cycles to ensure complete and homogeneous reactions.

The powder was subsequently pressed into polycrystalline rods (approximately 12 cm in length and 6 mm in diameter) under a hydrostatic pressure of 70 MPa for 30 minutes. These rods were then sintered at 1673 K for 12 hours in air. Single crystals were cultivated utilizing a vertical optical-image floating-zone furnace (Model HKZ,

SciDre) at Fudan University. The rod was initially traversed through the growth zone at a rapid speed of 20 mm h$^{-1}$ and 10-bar oxygen pressure to enhance density; then, the growth was carried out at a growth rate of 2 mm h$^{-1}$ within an oxygen pressure range of 130-150 bar, using a 5-kW xenon arc lamp as the illumination source.

**Calculation of superconducting volume fraction**

Following the approach outlined in Ref. [17], we calculate the superconducting volume fraction of Pr$_4$Ni$_3$O$_{10}$. The sample was a flat cylinder with a diameter of 210 μm and a thickness of 25 μm. The demagnetizing factor $N$ for this cylinder with an axial magnetic field was approximately 0.84 (Ref. [43]).

Using the refined lattice parameters obtained from XRD results, we compute the sample's volume at high pressure. The magnetic susceptibility in SI units was then calculated with the following equation:

$$\chi_{SI} = 4\pi\chi_{CGS} = \frac{4\pi\mu}{HV}$$

where $\mu$ denotes the measured magnetic moment at 5 K (after subtracting the constant background signal at the onset of $T_c$), and $V$ represents the cell volume for an $I4/mmm$ primitive cell with $Z = 2$ at high pressure.

After correcting for the demagnetizing factor, the susceptibility $\chi$ is approximately -0.85 at 40.2 GPa and -0.88 at 49.5 GPa for sample S3, corresponding to superconducting volume fractions of 85% and 88%, respectively.

**In situ high pressure synchrotron X-ray diffraction measurement**

The high-pressure XRD measurements were conducted at the beamline BL17UM of the Shanghai Synchrotron Radiation Facility with a wavelength of 0.4834 Å. The experiments utilized a symmetric DAC with 300 μm culet-sized anvils and rhenium gaskets. Helium was used as the pressure-transmitting medium to ensure optimal hydrostatic conditions.

*Note added*:

Recently, three independent studies reported signatures of superconductivity in Pr$_4$Ni$_3$O$_{10}$[44-46]. However, none of these studies observed zero resistance, and the onset

$T_c$ values are also lower than those observed in our samples.

† These authors contribute equally to this work.

* Correspondence and requests for materials should be addressed to J.Z. (zhaoj@fudan.edu.cn) or Q.S.Z.(zengqs@hpstar.ac.cn)

**Acknowledgments** We thank Y. Q. Hao for helpful discussions and W. W. Wang for assistance with the synchrotron XRD experiments. This work was supported by the National Key R&D Program of China (Grant No. 2022YFA1403202), the Key Program of the National Natural Science Foundation of China (Grant No. 12234006), the Innovation Program for Quantum Science and Technology (Grant No. 2024ZD0300103), and the Shanghai Municipal Science and Technology Major Project (Grant No. 2019SHZDZX01). Q.S.Z. and D.P. acknowledge support from Shanghai Key Laboratory of Material Frontiers Research in Extreme Environments, China (No. 22dz2260800) and the Shanghai Science and Technology Committee, China (No. 22JC1410300). Y.H.Z. acknowledges support from the Youth Foundation of the National Natural Science Foundation of China (Grant No. 12304173).

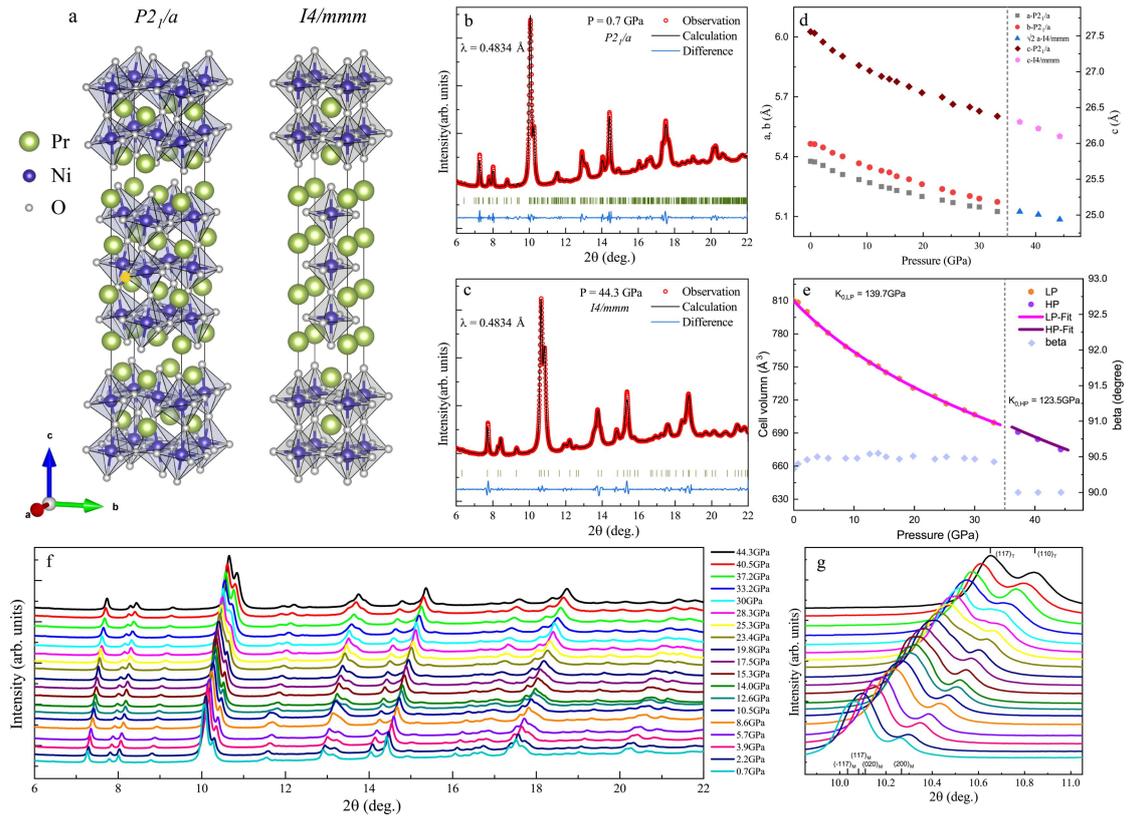

**Figure 1 | a**, Crystal structure of $Pr_4Ni_3O_{10}$ at ambient and low pressure (< 35 GPa, left) and high pressure (>35 GPa, right). **b-c**, Rietveld refinements of synchrotron-based powder XRD pattern of $Pr_4Ni_3O_{10}$ at 0.7 GPa (**b**) and 44.3 GPa (**c**) at room temperature, respectively. **d**, Pressure dependent lattice constants *a*, *b* and *c* determined from the Rietveld refinements. **e**, Pressure dependence of the cell volume and β angle obtained from Rietveld refinements, with the solid line representing a fit to the third-order Birch–Murnaghan equation of state, illustrating the compression behavior of the unit cell. **f**, Overall view of room-temperature synchrotron XRD patterns of pressurized $Pr_4Ni_3O_{10}$ ranging from 0.7 GPa to 44.3 GPa with a wavelength *λ* of 0.4834 Å. **g**, An enlarge view of diffraction peaks within the range of 9.85° ≤ 2θ ≤ 11.05°, illustrating the merging of the monoclinic $(0\ 2\ 0)_M$ and $(2\ 0\ 0)_M$ peaks into the tetragonal $(1\ 1\ 0)_T$ peak, indicating a structural phase transition from monoclinic to tetragonal.

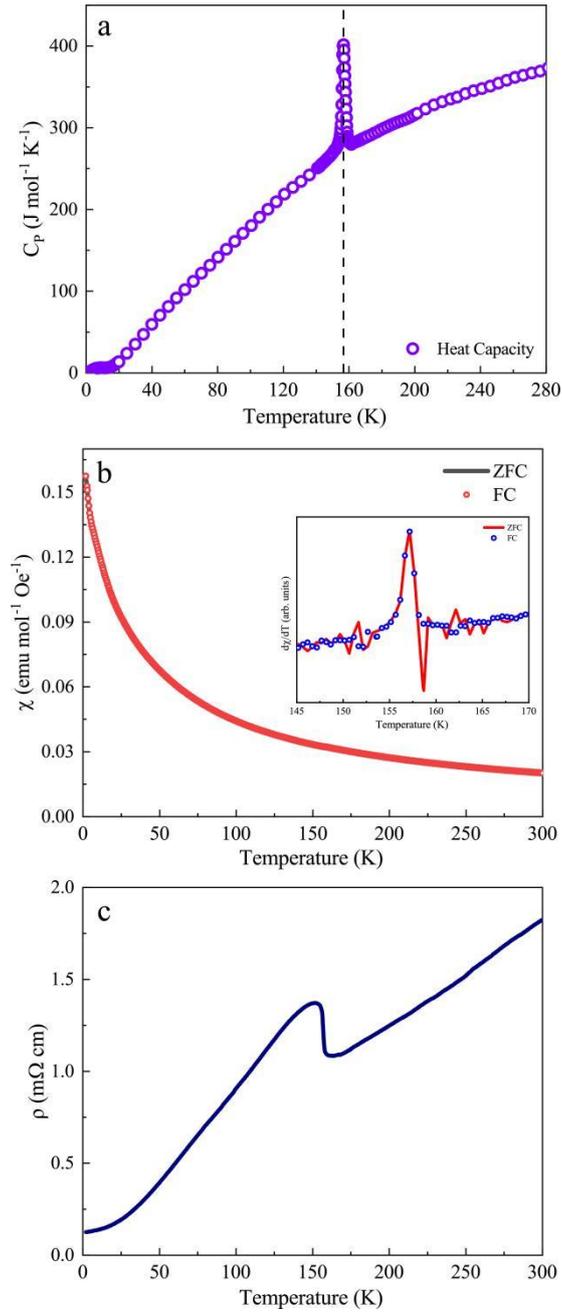

**Figure 2 | a**, Specific heat ($C_p$) of a $Pr_4Ni_3O_{10}$ single crystal measured from 1.8 K to 280 K. The density-wave order transition characterized by a sharp peak in the $C_p$ curve occurs at $T_d$ = 157 K. **b**, Magnetic susceptibility ($\chi$) of the $Pr_4Ni_3O_{10}$ single crystal measured from 1.8 K to 300 K with an applied field of 0.5 T, parallel to the *ab* plane. The inset shows the derivative of susceptibility, d$\chi$/dT. **c**, Resistance profile of the $Pr_4Ni_3O_{10}$ single crystal in the *ab* plane at ambient pressure.

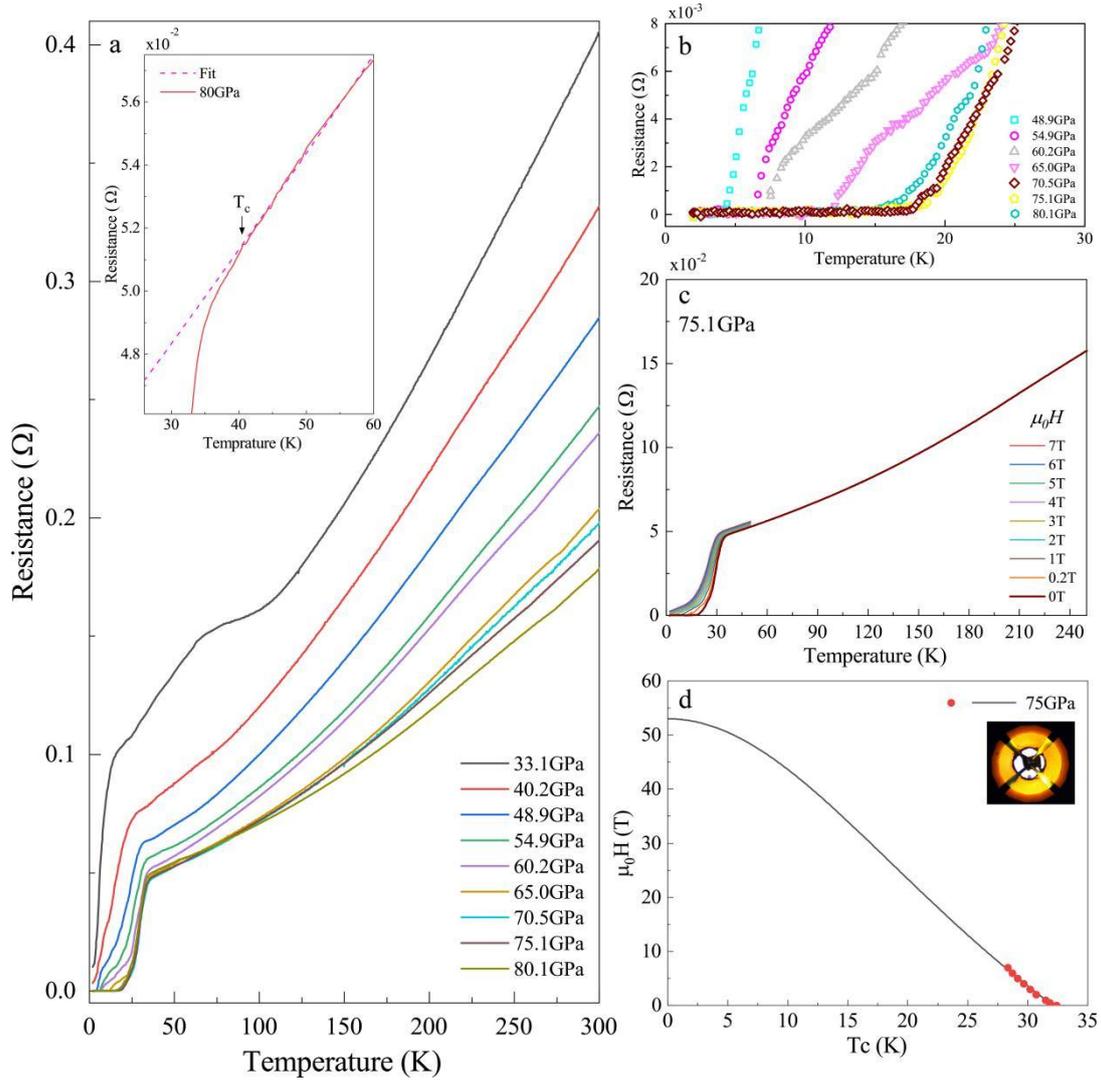

**Figure 3 | a**, Resistances of the $Pr_4Ni_3O_{10}$ single crystal at pressures ranging from 33.1 GPa to 80.1 GPa in the helium DAC for sample S1. The inset shows the onset $T_c$ = 40.5 K, which is defined as the temperature at which the *R-T* curve deviates from a power-law $\rho(T) = \rho_0 + AT^n$ fit. **b**, An enlarged view of the resistance curve near zero resistivity. **c**, Magnetic field effects on the superconducting transition in $Pr_4Ni_3O_{10}$ at 75.1 GPa. The magnetic fields are applied perpendicular to the *ab* plane. **d**, The Ginzburg-Landau fittings of the upper critical fields at 75.1 GPa. Inset shows a photograph of the electrodes used for high-pressure resistance measurements.

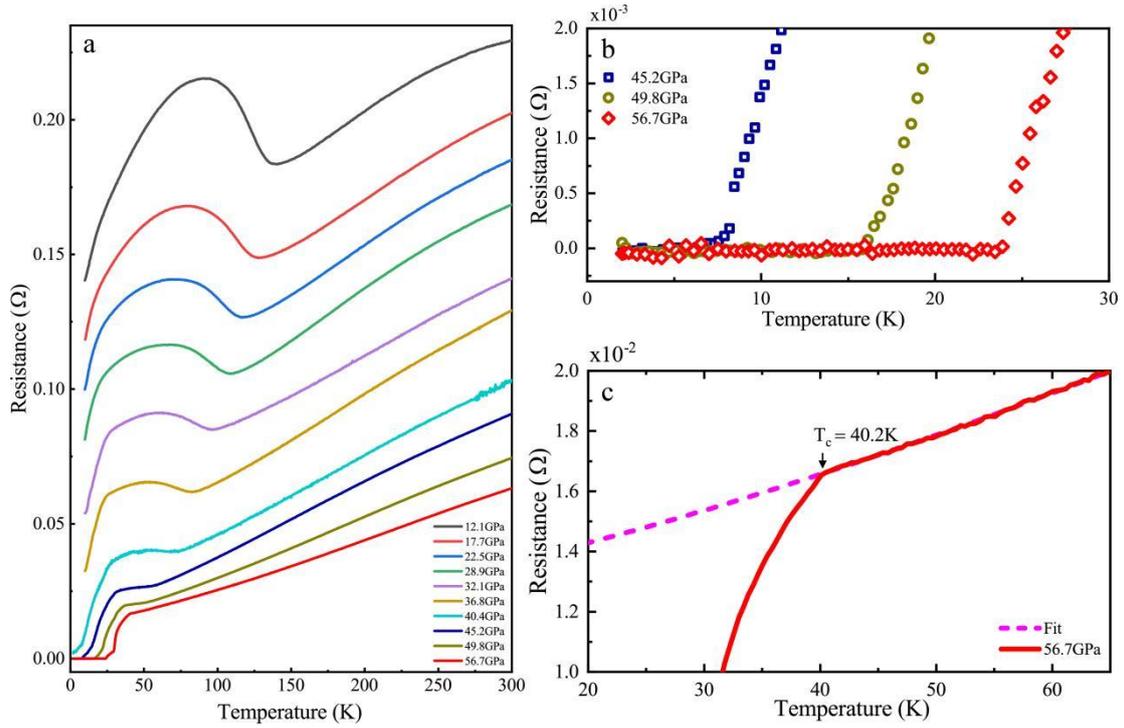

**Figure 4 | a,** Resistances of the $Pr_4Ni_3O_{10}$ single crystal sample S4 at pressures ranging from 12.1 GPa to 56.7 GPa in a helium DAC. The sample was carefully loaded to optimize hydrostaticity, allowing for a clear observation of the evolution of the density-wave transition and superconducting transition. **b**, An enlarged view of the resistance curve near zero resistance. **c**, An enlarged view of the resistance of sample S4 at 56.7 GPa near the onset $T_c$. The dashed line represents a fit using the $\rho(T) = \rho_0 + AT^n$ equation, indicating an onset $T_c$ of 40.2 K.

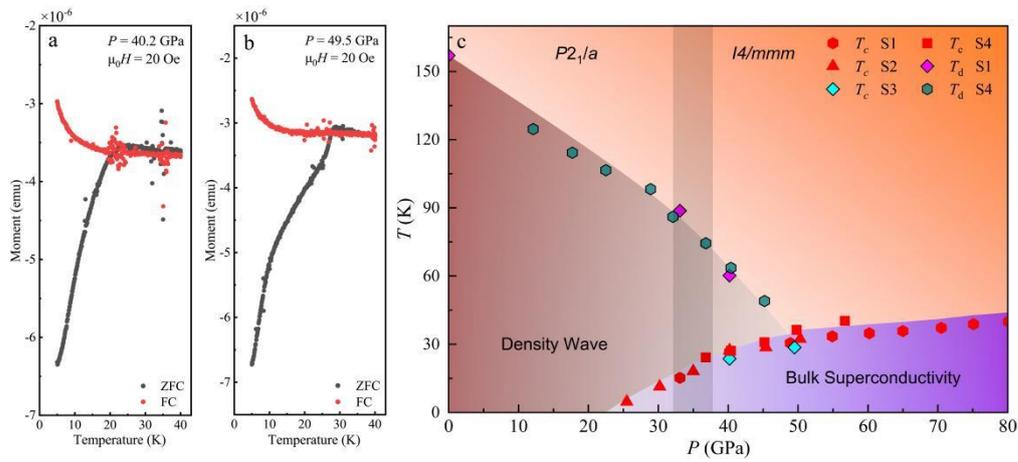

**Figure 5** | **a-b**, Temperature-dependendent DC susceptibilities of $Pr_4Ni_3O_{10}$ single crystals under various pressures. Both zero-field-cool (ZFC) and field-cool (FC) measurement modes were applied. Distinct superconducting diamagnetic responses below $T_c$ are observed in ZFC curves for 40.2 GPa (**a**), 49.5 GPa (**b**). **c**, Phase diagram of $Pr_4Ni_3O_{10}$ under pressure. The red solid hexagons, triangles and squares represent the $T_{c\_onset}$ of samples S1, S2 and S4 of resistance curve in the helium DAC, respectively. The blue diamonds and teal hexagon denote the density-wave transition $T_d$ of S1 and S4 determined from the anomaly in resistance curve. The turquoise diamonds denote the $T_c$ of samples S3 defined by magnetic susceptibility in the neon DAC.

**Reference:**

[1]. Lee, P. A., Nagaosa, N., and Wen, X.-G. Doping a Mott insulator: Physics of high-temperature superconductivity. *Reviews of modern physics*, 78 17-85 (2006).

[2]. Scalapino, D. J. A common thread: The pairing interaction for unconventional superconductors. *Reviews of modern physics*, 84 1383-1417 (2012).

[3]. Kamihara, Y., *et al.* Iron-based layered superconductor La[$O_{1-x}F_x$]FeAs (x= 0.05-0.12) with Tc= 26 K. *Journal of the American Chemical Society*, 130 3296-3297 (2008).

[4]. JG, B. Possible high-Tc super-conductivity in the Ba-La-Cu-O system. *Z Physik B*, 64 189-193 (1986).

[5]. Maeno, Y., *et al.* Superconductivity in a layered perovskite without copper. *Nature*, 372 532-534 (1994).

[6]. Wang, F. and Senthil, T. Twisted Hubbard model for Sr2IrO4: magnetism and possible high temperature superconductivity. *Physical Review Letters*, 106 136402 (2011).

[7]. Anisimov, V., Bukhvalov, D., and Rice, T. Electronic structure of possible nickelate analogs to the cuprates. *Physical Review B*, 59 7901 (1999).

[8]. Li, D., *et al.* Superconductivity in an infinite-layer nickelate. *Nature*, 572 624-627 (2019).

[9]. Sun, H., *et al.* Signatures of superconductivity near 80K in a nickelate under high pressure. *Nature*, 621 493-498 (2023).

[10]. Zhang, Y., *et al.* High-temperature superconductivity with zero resistance and strange-metal behaviour in $La_3Ni_2O_{7-\delta}$. *Nature Physics*, 20 1269-1273 (2024).

[11]. Wang, N., *et al.* Bulk high-temperature superconductivity in pressurized tetragonal $La_2PrNi_2O_7$. *Nature*, 634 579-584 (2024).

[12]. Wang, M., *et al.* Normal and Superconducting Properties of $La_3Ni_2O_7$. *Chinese Physics Letters*, 41 077402 (2024).

[13]. Zhou, Y., *et al.* Evidence of filamentary superconductivity in pressurized $La_3Ni_2O_7$ single crystals. *arXiv:2311.12361* (2023).

[14]. Wang, G., *et al.* Pressure-Induced Superconductivity In Polycrystalline $La_3Ni_2O_{7-\delta}$. *Physical Review X*, 14 011040 (2024).

[15]. Dong, Z., *et al.* Visualization of oxygen vacancies and self-doped ligand holes in $La_3Ni_2O_{7-\delta}$. *Nature*, 630 847-852 (2024).

[16]. Zhou, X., *et al.* Revealing nanoscale structural phase separation in $La_3Ni_2O_7$ single crystal via scanning near-field optical microscopy. *arXiv:2410.06602* (2024).

[17]. Zhu, Y., *et al.* Superconductivity in pressurized trilayer $La_4Ni_3O_{10-\delta}$ single crystals. *Nature*, 631 531-536 (2024).

[18]. Scott, B., *et al.* Layer dependence of the superconducting transition temperature of HgBa$_2$Ca$_{n-1}$Cu$_n$O$_{2n+2+\delta}$. *Physica C: Superconductivity*, 230 239-245 (1994).

[19]. Kuzemskaya, I., Kuzemsky, A., and Cheglokov, A. Superconducting properties of the family of mercurocuprates and role of layered structure. *Journal of low temperature physics*, 118 147-152 (2000).

[20]. Iyo, A., *et al.* T-$_c$ vs n Relationship for Multilayered High-T$_c$ Superconductors. *Journal of the Physical Society of Japan*, 76 094711 (2007).

[21]. Li, Q., *et al.* Signature of superconductivity in pressurized La$_4$Ni$_3$O$_{10}$. *Chinese Physics Letters*, 41 017401 (2024).

[22]. Zhang, M., *et al.* Superconductivity in trilayer nickelate La$_4$Ni$_3$O$_{10}$ under pressure. *arXiv:2311.07423* (2023).

[23]. Sakakibara, H., *et al.* Theoretical analysis on the possibility of superconductivity in the trilayer Ruddlesden-Popper nickelate La$_4$Ni$_3$O$_{10}$ under pressure and its experimental examination: Comparison with La$_3$Ni$_2$O$_7$. *Physical Review B*, 109 144511 (2024).

[24]. Zhang, Y., *et al.* Trends in electronic structures and s±-wave pairing for the rare-earth series in bilayer nickelate superconductor R$_3$Ni$_2$O$_7$. *Physical Review B*, 108 165141 (2023).

[25]. Pan, Z., *et al.* Effect of rare-earth element substitution in superconducting R$_3$Ni$_2$O$_7$ under pressure. *Chinese Physics Letters*, 41 087401 (2024).

[26]. Zhang, J., *et al.* High oxygen pressure floating zone growth and crystal structure of the metallic nickelates R$_4$Ni$_3$O$_{10}$ (R= La, Pr). *Physical Review Materials*, 4 083402 (2020).

[27]. Song, J., *et al.* Structure, electrical conductivity and oxygen transport properties of Ruddlesden-Popper phases Ln$_{n+1}$Ni$_n$O$_{3n+1}$ (Ln= La, Pr and Nd; n= 1, 2 and 3). *Journal of Materials Chemistry A*, 8 22206-22221 (2020).

[28]. Samarakoon, A. M., *et al.* Bootstrapped Dimensional Crossover of a Spin Density Wave. *Physical Review X*, 13 041018 (2023).

[29]. Shen, Y., *et al.* Structural and magnetic phase diagram of CrAs and its relationship with pressure-induced superconductivity. *Physical Review B*, 93 060503 (2016).

[30]. Stewart, G. Superconductivity in iron compounds. *Reviews of modern physics*, 83 1589-1652 (2011).

[31]. Qin, Q., Wang, J., and Yang, Y.-f. Frustrated Superconductivity in the Trilayer Nickelate La4Ni3O10. *The Innovation Materials*, 2 100102 (2024).

[32]. LaBollita, H., *et al.* Electronic structure and magnetic tendencies of trilayer La$_4$Ni$_3$O$_{10}$ under pressure: Structural transition, molecular orbitals, and layer differentiation. *Physical Review B*, 109 195151 (2024).

[33]. Wang, J.-X., *et al.* Non-Fermi liquid and Hund correlation in $La_4Ni_3O_{10}$ under high pressure. *Physical Review B*, 109 165140 (2024).

[34]. Huo, Z., *et al.* Electronic Correlations and Hund's Rule Coupling in Trilayer Nickelate $La_4Ni_3O_{10}$. *arXiv:2407.00327* (2024).

[35]. Chen, C.-Q., *et al.* Trilayer multiorbital models of $La_4Ni_3O_{10}$. *Physical Review B*, 110 014503 (2024).

[36]. Zhang, Y., *et al.* Prediction of s±-wave superconductivity enhanced by electronic doping in trilayer nickelates $La_4Ni_3O_{10}$ under pressure. *Physical Review Letters*, 133 136001 (2024).

[37]. Oh, H., Zhou, B., and Zhang, Y.-H. Type II tJ model in charge transfer regime in bilayer $La_3Ni_2O_7$ and trilayer $La_4Ni_3O_{10}$. *arXiv:2405.00092* (2024).

[38]. Leonov, I. Electronic structure and magnetic correlations in the trilayer nickelate superconductor $La_4Ni_3O_{10}$ under pressure. *Physical Review B*, 109 235123 (2024).

[39]. Huang, J. and Zhou, T. Interlayer pairing-induced partially gapped Fermi surface in trilayer $La_4Ni_3O_{10}$ superconductors. *Physical Review B*, 110 L060506 (2024).

[40]. Yang, Q.-G., *et al.* Effective model and s±-wave superconductivity in trilayer nickelate $La_4Ni_3O_{10}$. *Physical Review B*, 109 L220506 (2024).

[41]. Zhou, G., *et al.* Ambient-pressure superconductivity onset above 40 K in bilayer nickelate ultrathin films. *arXiv:2412.16622* (2024).

[42]. Ko, E. K., *et al.* Signatures of ambient pressure superconductivity in thin film $La_3Ni_2O_7$. *Nature* https://doi.org/10.1038/s41586-41024-08525-41583 (2024).

[43]. Prozorov, R. and Kogan, V. G. Effective demagnetizing factors of diamagnetic samples of various shapes. *Physical review applied*, 10 014030 (2018).

[44]. Huang, X., *et al.* Signature of Superconductivity in Pressurized Trilayer-nickelate $Pr_4Ni_3O_{10-\delta}$. *arXiv:2410.07861* (2024).

[45]. Chen, X., *et al.* Non-bulk Superconductivity in $Pr_4Ni_3O_{10}$ Single Crystals Under Pressure. *arXiv:2410.10666* (2024).

[46]. Pei, C., *et al.* Pressure-Induced Superconductivity in $Pr_4Ni_3O_{10}$ Single Crystals. *arXiv:2411.08677* (2024).

# Supplementary Information for

# Bulk superconductivity in pressurized trilayer nickelate Pr$_4$Ni$_3$O$_{10}$ single crystals

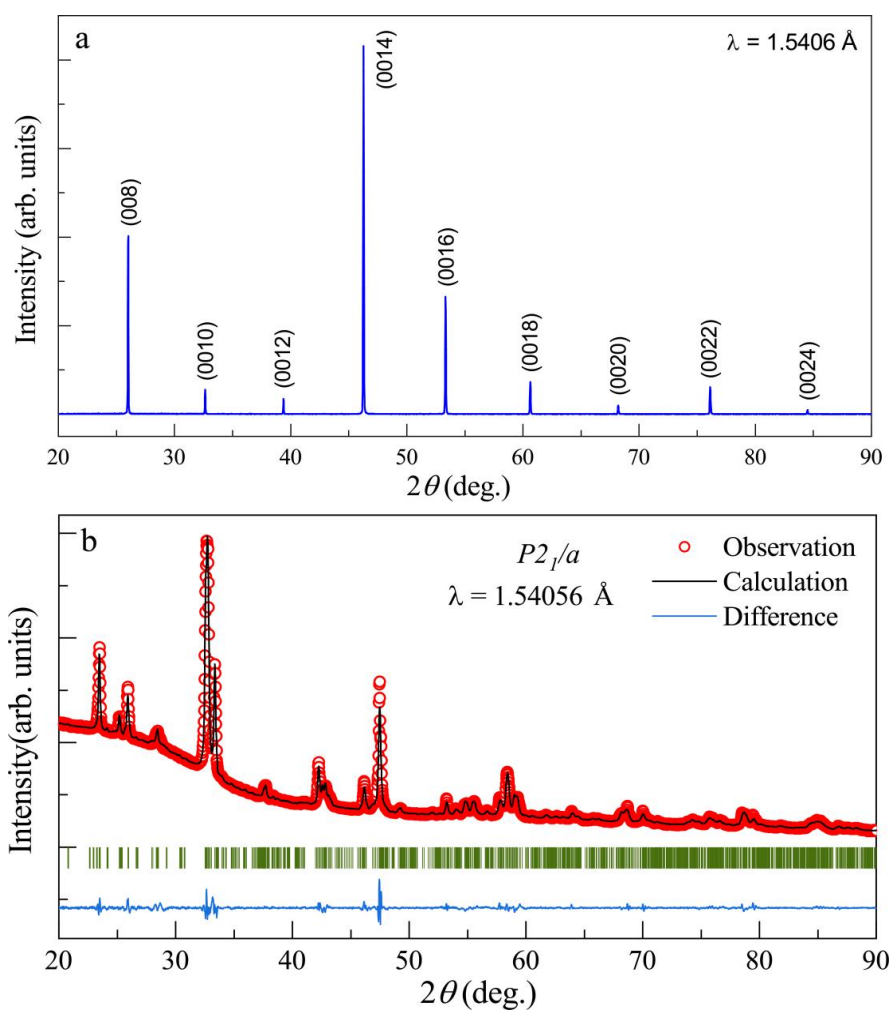

**Supplementary Fig. 1 | a**, X-ray diffraction (XRD) measurements of a naturally cleaved Pr$_4$Ni$_3$O$_{10}$ single crystal along the crystallographic *ab* plane, which is consistent with a trilayer structure with no detectable impurity phases. **b**, Rietveld refinement of the XRD pattern of powdered Pr$_4$Ni$_3$O$_{10}$ single crystal at ambient pressure and room temperature. The refined data, which fits well with the *P*2$_1$/*a* space group, confirms the phase purity of the sample. The measurements were conducted using a Bruker D8 Discover diffractometer with Cu Kα radiation.

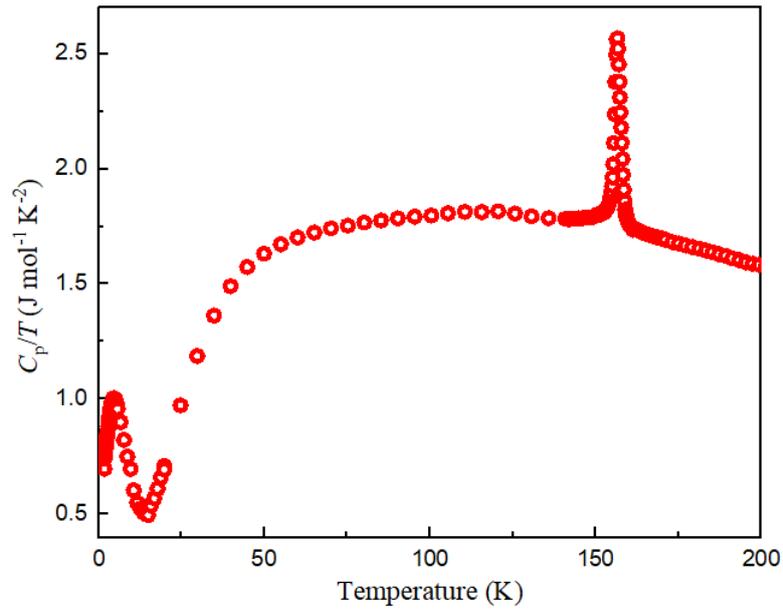

**Supplementary Fig. 2** The $C_p/T$ data of $Pr_4Ni_3O_{10}$ ranging from 1.8 K to 200 K at ambient pressure. An anomaly near 5 K is observed, which is associated with the magnetic ordering of the Pr moments. The prominent peak around 157 K marks the transition associated with spin/charge density wave ordering.

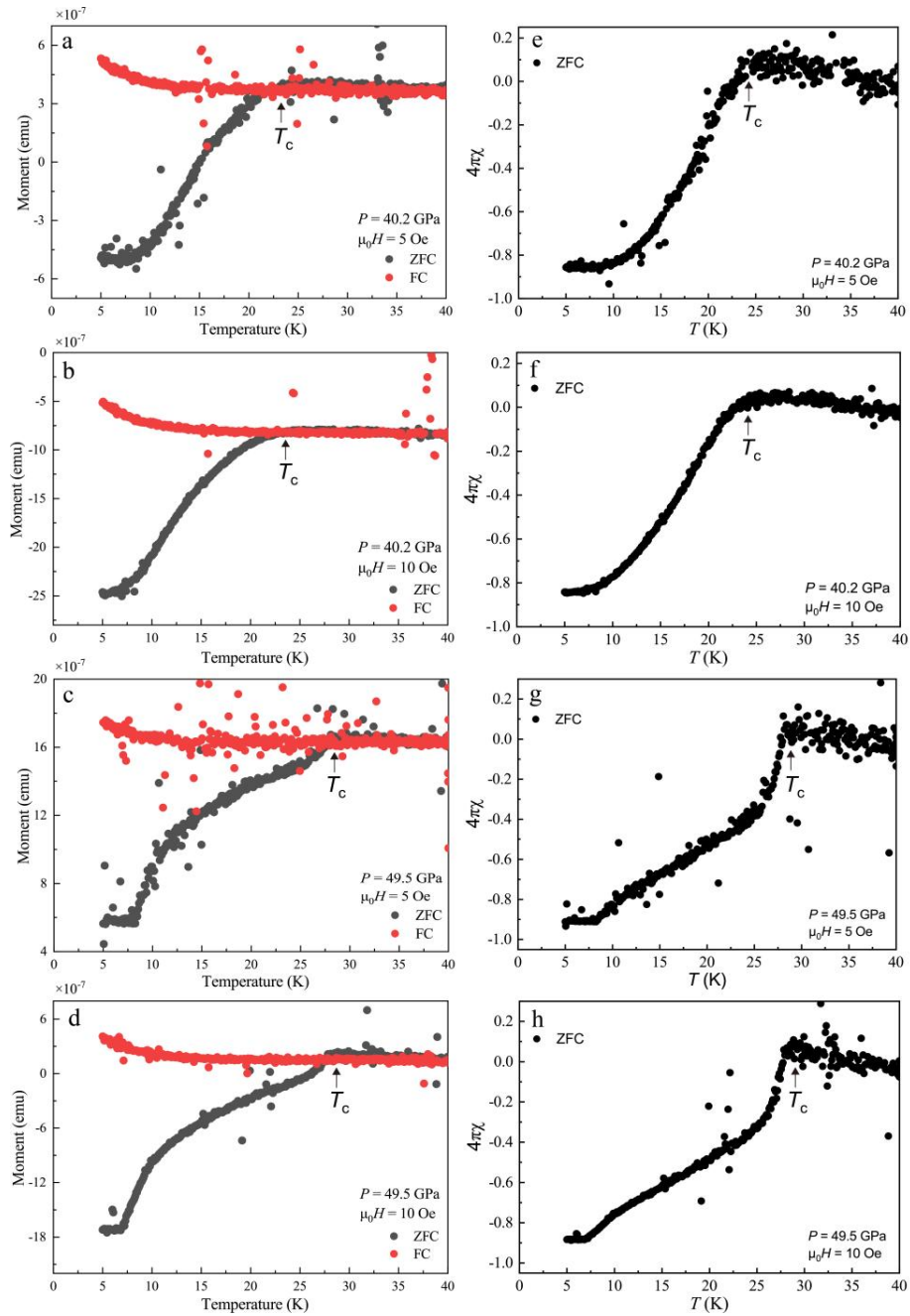

**Supplementary Fig. 3 | a-d**, Raw temperature dependent DC susceptibilities of $Pr_4Ni_3O_{10}$ single crystals under various pressures and magnetic fields of sample 3 (S3). Both zero-field-cool (ZFC) and field-cool (FC) measurement modes were applied. The magnetic fields are applied along perpendicular to the *ab* plane. Distinct superconducting diamagnetic responses below $T_c$ are observed in ZFC curves for 40.2 GPa, 5 Oe (**a**), 40.2 GPa, 10 Oe (**b**), 49.5 GPa 5 Oe (**c**) and 49.5 GPa 10 Oe (**d**). In the normal state, the background signals are primarily due to the pressure cell and rhenium gaskets. **e-f**, Superconducting volume fractions calculated from the DC susceptibility measurements at indicated pressure and field using the method described in the main text.

**Supplementary Table 1** | Refined parameters of pressurized $Pr_4Ni_3O_{10}$ obtained from the Rietveld refinements of XRD data.

| Pressure (GPa) | a(Å) | b(Å) | c(Å) | β(°) | GOF | $R_p$ (%) | $R_{wp}$ (%) | Space group |
|---|---|---|---|---|---|---|---|---|
| 0 | 5.3773 | 5.4652 | 27.5585 | 90.33 | 3.60 | 16.2 | 13.6 | *$P2_1/a(Z=4)$* |
| 0.7 | 5.3739 | 5.4637 | 27.5374 | 90.40 | 0.65 | 5.06 | 5.22 | *$P2_1/a(Z=4)$* |
| 2.2 | 5.3569 | 5.4470 | 27.4138 | 90.46 | 1.00 | 12.2 | 11.7 | *$P2_1/a(Z=4)$* |
| 3.9 | 5.3306 | 5.4207 | 27.2996 | 90.50 | 1.10 | 11.0 | 11.2 | *$P2_1/a(Z=4)$* |
| 5.7 | 5.3114 | 5.4022 | 27.2158 | 90.48 | 0.97 | 9.94 | 9.99 | *$P2_1/a(Z=4)$* |
| 8.6 | 5.2860 | 5.3678 | 27.0867 | 90.48 | 1.00 | 10.3 | 10.8 | *$P2_1/a(Z=4)$* |
| 10.5 | 5.2702 | 5.3477 | 27.0132 | 90.48 | 1.30 | 14.7 | 14.3 | *$P2_1/a(Z=4)$* |
| 12.6 | 5.2508 | 5.3306 | 26.9346 | 90.54 | 0.88 | 9.37 | 9.11 | *$P2_1/a(Z=4)$* |
| 14.0 | 5.2430 | 5.3224 | 26.8974 | 90.55 | 1.10 | 10.0 | 10.6 | *$P2_1/a(Z=4)$* |
| 15.3 | 5.2312 | 5.3028 | 26.8602 | 90.50 | 0.89 | 9.20 | 8.94 | *$P2_1/a(Z=4)$* |
| 17.5 | 5.2194 | 5.2877 | 26.7915 | 90.47 | 1.00 | 10.8 | 10.1 | *$P2_1/a(Z=4)$* |
| 19.8 | 5.2011 | 5.2631 | 26.7059 | 90.51 | 1.00 | 8.98 | 9.83 | *$P2_1/a(Z=4)$* |
| 23.4 | 5.1830 | 5.2387 | 26.6444 | 90.47 | 0.95 | 8.94 | 9.47 | *$P2_1/a(Z=4)$* |
| 25.3 | 5.1704 | 5.2221 | 26.5460 | 90.50 | 1.40 | 10.9 | 12.5 | *$P2_1/a(Z=4)$* |
| 28.3 | 5.1533 | 5.2035 | 26.5066 | 90.48 | 0.96 | 9.66 | 9.13 | *$P2_1/a(Z=4)$* |
| 30.0 | 5.1481 | 5.1909 | 26.4494 | 90.47 | 1.10 | 10.3 | 10.3 | *$P2_1/a(Z=4)$* |
| 33.2 | 5.1254 | 5.1743 | 26.3784 | 90.43 | 1.30 | 13.8 | 12.4 | *$P2_1/a(Z=4)$* |
| 37.2 | 3.6246 | 3.6246 | 26.2991 | 90 | 1.10 | 12.1 | 11.3 | *I4/mmm* |
| 40.5 | 3.6138 | 3.6138 | 26.2060 | 90 | 0.87 | 11.7 | 10.3 | *I4/mmm* |
| 44.3 | 3.5969 | 3.5969 | 26.0985 | 90 | 0.64 | 6.98 | 7.00 | *I4/mmm* |